\documentstyle[aps,epsfig,multicol,amssymb,amsfonts]{revtex}
\tighten
\renewcommand{\narrowtext} 
{\begin{multicols}{2}\global\columnwidth21pc} 
\renewcommand{\widetext} 
{\end{multicols}\global\columnwidth42.5pc} 
\newcommand{\be}{\begin{eqnarray}}
\newcommand{\ee}{\end{eqnarray}}
\def\beq{\begin{equation}}
\def\eeq{\end{equation}}
\begin{document} 
\draft 
\title{Effect of a magnetic flux on the critical behavior of a system with long range hopping} 
\author{Antonio M. Garc\'{\i}a-Garc\'{\i}a} 
\address{Laboratoire de Physique Th\'eorique et Mod\`eles 
Statistiques, B\^at. 100, \\ Universit\'e de Paris-Sud, 
91405 Orsay Cedex, France}
\maketitle
\begin{abstract}
 We study the effect of a magnetic flux
 in a 1D disordered wire with long range hopping.
 It is shown that this model  
 is  at the metal-insulator transition (MIT) for 
 all disorder values and the spectral correlations
 are given by critical statistics. In the 
 weak disorder regime a smooth transition between 
 orthogonal and unitary symmetry is observed 
 as the flux strength increases. By contrast, in the strong disorder regime  
  the spectral correlations are almost 
 flux independent. It is also conjectured that the two  
level correlation function for arbitrary flux is 
given by the dynamical density-density correlations
 of the Calogero-Sutherland (CS) model at finite temperature. 
Finally we describe the classical dynamics of the model 
and its relevance to quantum chaos.            
\end{abstract}

\pacs{PACS numbers: 72.15.Rn, 71.30.+h, 05.45.Df, 05.40.-a} 
\narrowtext
The addition of disorder to an otherwise metallic sample strongly modifies its properties. 
As disorder increases eigenfunctions start to localize. A 
MIT is observed in the thermodynamic
 limit in systems of dimension greater than two.
 The moments of the 
 eigenfunctions at the MIT show anomalous 
scaling with the sample size \cite{wegner80} $L$, 
$P_q=\int d^dr |\psi({\bf r})|^{2q}\propto L^{-D_q(q-1)}$,
$P_q$ is the inverse participation ratio 
and $D_q$ is a set of exponents describing the transition.
 Thus the scaling at the MIT is 
  in between that of a good metal, 
$P_q\propto L^{-d(q-1)}$, and that of an
insulator (localized  eigenfunctions),  $P_q\propto L^0$. 
 Eigenfunctions with such anomalous scaling
 are named ``multifractals'', for a review see \cite{janssen}. Signatures of a MIT
  are found not only in the eigenfunctions \cite{castpel} 
but also in the spectral fluctuations \cite{sko}. 
  It was argued in \cite{chi} that 
 although level repulsion typical of the metallic regime is still present at the MIT, 
 the long range correlations are weaken due to the multifractal 
 character of the wavefunctions.   
 The  number variance was claimed to 
 be asymptotically linear with a slope 
 around $0.25$\cite{chi}. These features, level repulsion and sub-Poisson number 
 variance combined with the scale 
 invariance of the spectrum \cite{sko} are named 'critical statistics'\cite{kravtsov97}
 and are considered genuine spectral signatures of a MIT.    
Eigenfunction 
and spectral properties 
  are indeed related, for multifractal eigenstates not 
 too sparse the slope of the number variance $\chi $ is related to the multifractal 
exponent $D_2$ by $\chi=\frac{d-D_{2}}{2d}$ \cite{chalker96},
where $d$ is the dimension of the space.
Different generalized random matrix model (gRMM) have been successfully 
employed to describe 'critical statistics'\cite{Moshe}.
 Recently \cite{ant3,tsv}, the spectral correlations of one of those gRMM 
 was demonstrated to be equivalent to the spatial correlations of 
the Calogero-Sutherland (CS) model \cite{calo} at finite 
 temperature. Temperature in the CS model is  
related with the multifractal exponent $D_2$ at the transition.   

 Short range Anderson \cite{anderson} models have
 been broadly utilized for numerical investigation at MIT.   
However in certain systems, like glasses with strong dipole 
interactions \cite{levitov90}, long range hopping is possible and thus 
the Anderson model must be modified accordingly. 
  Although the introduction of long range hopping dates back to the 
 famous Anderson's paper on localization\cite{anderson},  
 these models did not attract too much attention until the numerical work of Oono \cite{ono,yu} 
and the 
 renormalization group treatment of Levitov \cite{levitov90} in the context of glassy systems. 
The main conclusion of these works was that power low 
 hopping may induce a MIT even in one dimensional systems if the exponent of the 
 hopping decay matched the dimension of the space.   
 A related problem, a random banded matrix 
with a power law decay was discussed in \cite{prbm}. It was shown analytically that  
 for a $1/r$ band decay the eigenstates are multifractal 
 and the spectral correlations resemble those at the MIT. 
 Intense numerical study in recent years \cite{ever} has corroborated
 the  close relation between this random banded matrix 
model and the Anderson model at the MIT.
      
 Another related issue of current interest is the effect of a 
magnetic field at the MIT \cite{hof,bat}. In the metallic regime, an exact 
 analytical treatment 
 was developed in \cite{efetov}. It turns out that, in this limit, the
 two level correlation function describing the 
 crossover between orthogonal (no flux) and unitary symmetry is equivalent to 
the dynamical density-density  correlations of the 
 CS model \cite{shu} (see also \cite{ha} for different initial conditions) provided that ``time'' in the CS model is traded by magnetic field in 
the disordered system. Numerical results at the 3D MIT \cite{hof} show that 
 the magnetic flux still influences the short-range spectral correlations. The impact 
  on long range correlations is still not settled \cite{bat} though there is agreement
 that the effect, if any, must be small.   
 
 One of the main aims of this paper is to further 
 clarify this issue by investigating a one dimensional (1D)
 wire with power low hopping and a magnetic flux attached to it. This model
 presents typical features of a MIT for all values of disorder with 
 the advantage that much larger volumes can be simulated. 
We also conjecture,
 based on the analogy with the CS model above mentioned,
 a exact expression for the two level spectral correlation function at the MIT for arbitrary 
 flux.   

   Our starting point is the following  
  Hamiltonian describing a 1D closed  
wire with long-range hopping,
 \begin{equation}
\label{mod}
\small{H =\sum_n \epsilon_n |n \rangle \langle n|+\sum_{n \neq m}
\epsilon_{nm}e^{-2i\theta_{nm}}F(r,b)|n \rangle \langle m|}
\end{equation}
where, $n,m$ label the lattice sites, $r=n-m$, 
 $\epsilon_n$ and $\epsilon_{nm}$ are two
 sets of random number both distributed in a box $[-W,W]$,
$F(r,b)$ is the long range hopping term, 
\be
F(r,b)=\left[1+{1\over \lambda^2}{\sin^2(\pi r/N)\over(\pi/N)^2}\right]^{-1/2}
\ee
 $0<\lambda<\infty$ is the band size and $N$ is the system size    
  Since $F(r,b) \sim 1/r$ for 
$N\gg r$ our model is critical, namely reproduce 
 typical features of a MIT  for all values of $\lambda$ \cite{prbm}.  
 Sine-like interaction is introduced in order to assure periodicity.
The phase factor  (Peierls substitution) above is related to the magnetic flux 
(in units of the fundamental flux $h/c$) by, 
$\theta_{nm}=\int_{n}^{m}\vec{ A}d\vec{r}=\alpha(n-m)/N$ for $n-m< N/2$ and 
$\theta_{nm}=\int_{n}^{m}\vec{A} d\vec{r}=
\alpha(1-(n-m)/N)$ for $n-m> N/2$, $\alpha$ is the magnetic flux.
 $\vec{A}$ is the (constant) potential vector. 
  Such Hamiltonian may be relevant for glassy systems \cite{yu}, 
  quantum effect of classical anomalous diffusion \cite{rev} and
 quantum chaos \cite{levitov97}.

The spectral fluctuations are studied by   
  direct diagonalization of the Hamiltonian (\ref{mod})
 for different sizes ranging from $N=500$ to $N=8000$. The eigenvalues thus obtained 
 are unfolded with respect to the mean spectral density.   
 The 
 number of different realizations of disorder is chosen such that
 for each $N$ the total number of eigenvalue be at least $2 \times 10^5$. In order 
 to avoid edge effects, only $30\%$ 
of the eigenvalues around the center of the band are utilized. Without loss of generality 
 we have set $W=1$.

First we investigate short range correlations by evaluating $P(s)$, 
the level spacing distribution
 (LSD). This correlator gives the probability of having two eigenvalues 
at a distance
 $s$. In the insulator regime (uncorrelated eigenvalues) $P(s)=e^{-s}$, 
 in the metallic regime (Wigner-Dyson statistics) 
$P(s)\sim s^\beta$ for  $s \ll 1$  
(the presence of a flux drives the 
 spectrum from $\beta=1$ orthogonal to $\beta=2$ unitary symmetry).
At criticality one expects level repulsion as in Wigner-Dyson 
 statistics $s \ll 1$ and exponential decay for $s \gg 1$, 
$P(s)\sim e^{-\kappa s}$ with $\kappa > 1$ \cite{nishi}. 
In figure 1 we plot $P(s)$ for different flux values 
$\alpha$ and $\lambda=2$. As the flux strength 
increases, a transition is clearly observed between 
 orthogonal and unitary symmetry. For $s \ll 1$ (see inset) level repulsion is 
 still present, $P(s)\sim s^{\beta}$ is Wigner-Dyson like 
 but with the prefactor modified by the flux. 
  The strong disorder regime is reached by choosing a coupling constant
 ($\lambda=0.44$)
 such that the resulting spectrum resembles the one at the 3D MIT.
 As observed in figure 2, 
  $P(s)$ is almost independent of the flux strength. For $s>1$ 
  the decay is exponential (see inset) and independent of the flux. For $s \ll 1$, 
  the effect of the flux is still important, only for $\alpha \sim 0$ the expected $P(s) \sim s$
 behavior for orthogonal symmetry is recovered.   
 Such result agrees
 with previous \cite{hof,bat} numerical simulations.   
 Let us now move to long range spectral 
 correlations.
\begin{figure}
\includegraphics[width=0.8\columnwidth,clip]{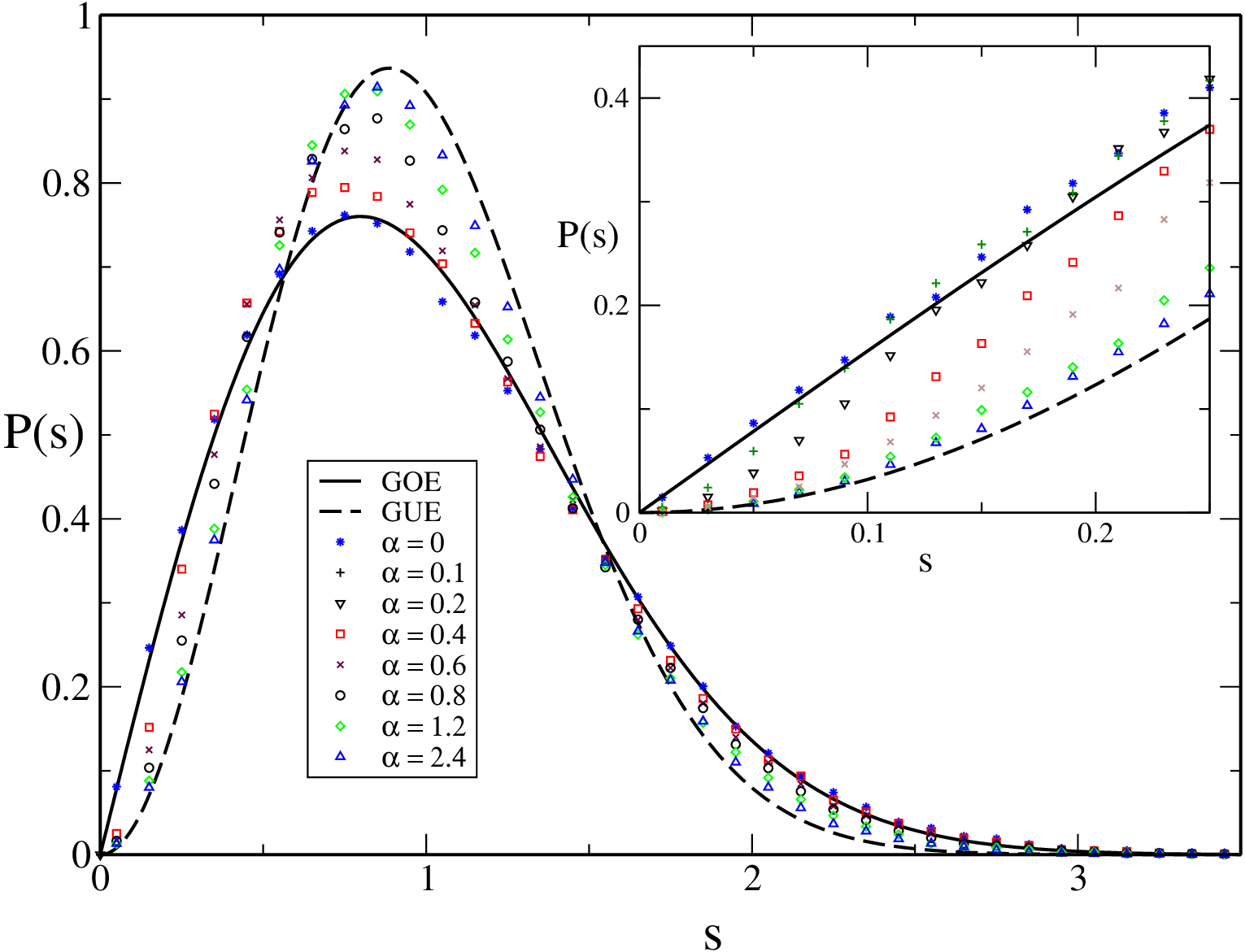}
\caption{Level spacing distribution $P(s)$ for $\lambda=2$ and different fluxes  
$\alpha$. 
A transition between the Gaussian Orthogonal ($GOE$) and Gaussian Unitary ($GUE$) 
 ensemble is observed as the flux is increased, even
 the asymptotic behavior of $P(s)$ is modified by the flux.}  
\label{fig1} 
\end{figure}          
The number variance $\Sigma^{2}(L)= \langle L \rangle^2 - \langle L^2 \rangle=L-
2\int_{0}^{L}ds(L-s)R_2(s,\alpha)$ ($R_2(s,\alpha)$ is the two level correlation function)
  measures
the stiffness of the spectrum. Fluctuations are small in the metallic
 regime (Wigner-Dyson statistics)
  with $\Sigma^{2}(L) \sim \log(L)$ for $L \gg 1$.
For localized eigenstates, the eigenvalues are
uncorrelated (Poisson statistics) and  $\Sigma^{2}(L)=L$.
As mentioned earlier, the number variance is asymptotically proportional
to $\chi L$ ($\chi < 1$) at the MIT.
In figure 3 we  plot the number variance in the weak disorder regime ($\lambda = 2$)
 for different flux values.  
 As the flux is increased, a smooth transition is  
 observed between critical $GOE$ \cite{tsv,ant3}  
 and critical $GUE$ \cite{Moshe}. We remark the slope is not modified
 by the flux. This is in apparent disagreement with the results for a 2D Anderson model 
 in the weak multifractality regime where
 the slope is two times smaller in the
 unitary case \cite{efetov}. The reason for that is the way in which the flux is introduced due
 to the long range hopping. 
If the flux in (\ref{mod}) is introduced as usual in short range models the 
 2D Anderson model results are recovered.    
 The asymptotically linear number variance  
 (see fig 3) together with the level repulsion indicates that   
  our model is described by critical statistics \cite{kravtsov97}.  
  The strong disorder regime is explored
  by  choosing a coupling constant ($\lambda=0.44$)
 such that the slope of the number variance 
$\chi \sim 0.29$  be similar to that at the 3D MIT. We observe (fig 4) 
 that the flux only modifies
 the spectral correlations up to ($L \sim 1$) (see inset fig 4). Beyond this point 
 the number variance is independent of the flux, thus suggesting that symmetry is 
 washed out by disorder. We want to stress that the transition from weak to strong disorder 
 is not sharp. As $\lambda$ decreases, the flux impact 
 decreases smoothly. Curiously, until values of $\lambda$ very close to those  
at the 3D MIT, the effect, although small, is still sizable. Such behavior
 makes more difficult an accurate account of the flux at the 3D MIT. Finally let us 
 mention that although not shown in the figures it was explicitly checked that the
 spectral correlators are not dependent on the size of the matrix as expected at a MIT.  
\begin{figure}
\includegraphics[width=0.8\columnwidth,clip]{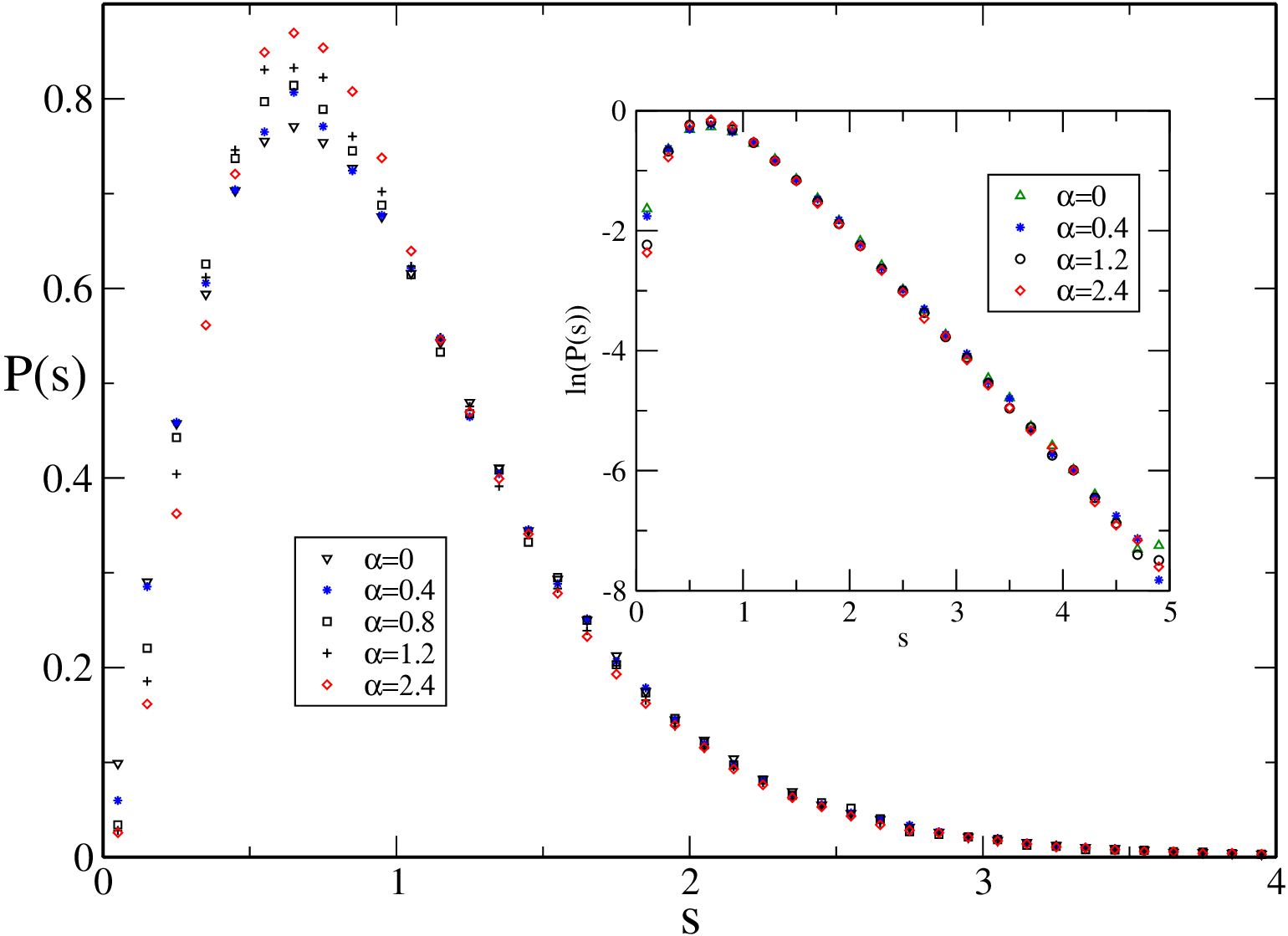}
\caption{P(s) for $\lambda=0.44$ and different fluxes. Only for $s \ll 1$ the system
 is sensitive to the flux. For $s \gg 1$ all curves become indistinguishable (inset).}  
\label{fig2} 
\end{figure}
 After discussing the numerical results we propose an exact analytical relation for  
 the two level correlation function of (\ref{mod}). We 
claim that such correlation function is equivalent to the dynamical density-density
 correlations of the CS model \cite{calo}, 
$\small{{\hat H_{CS}}=-\sum_{j}\frac{\partial^2}{\partial
x_{j}^{2}}+
\frac {\beta(\beta - 2)}4 \sum_{i \neq j}\frac{1}{(x_{i}-x_{j})^2}+
\frac{1}{4N^2}\sum_j x_{j}^{2}}$\\
 at finite temperature where the position
 of the CS particles corresponds with the (unfolded) eigenenergies of (\ref{mod}), 
``time'' $\sim \alpha^2$ and 
  temperature $h$ is related to $\lambda$. Let us first argue how these density-density 
correlations  
are obtained \cite{tsv,shu}. 
The density of probability of the CS ground state (zero temperature) for ($\beta=1,2,4$) is 
 equivalent to the joint distribution of eigenvalue of the Gaussian random 
matrix models (GRMM). In the 
 large $N$ limit, the two point spectral correlations of the GRMM
 are explicitly expressed through a spectral kernel.  Such kernel 
 in the language of the CS model corresponds with the amplitude of probability 
 of having two free fermions in a ensemble of $N$ around two arbitrary positions.      
 The idea is that,
 based on the Luttinger liquid nature of the CS model \cite{kawa},
 the density-density correlations of
the CS model at finite low temperature can be 
 obtained by using the results of GRMM but   
replacing the  spectral kernel above mentioned
by its finite
temperature  analogue. In the grand canonical ensemble \cite{Moshe},
$K(s=x-y,0)_h=\sum_n \frac{\psi_n(x) \psi^{\dagger}_n(y)}{1 + z^{-1} e^{E_n/h}}$
where $\psi_n$ are the single particle wave functions for free fermions and $E_n = n/N$.
In the large $N$ limit the wave functions are replaced by plane waves with energy given by
$k^2$. 

 Dynamical density-density correlations \cite{shu} can be included in this formalism
  by expressing the time dependent density
 as  $\rho(s,t\sim \alpha^2) = e^{{\hat H_{CS}}t}\rho(s)e^{-{{\hat H}_{CS}}t}$. The
  crossover between orthogonal and unitary symmetry in the GRMM \cite{efetov} 
  corresponds in the language of the CS model
  with density-density correlations 
  with initial conditions given by particles distributed according to
the  CS model 
 for $\beta = 1$ and then evolving for $t > 0$
 according to the CS model for $\beta = 2$ (free fermions). The density-density correlations
  in this case \cite{shu}
 are also expressed through a spectral kernel as the one mentioned above but now involving
 the propagation of free fermions (hole and particle). 
  Combining both effect, the dynamical 
(with the inital conditions above mentioned) density-density correlations
 of the CS model at finite temperature 
are  given  
by (see \cite{ant3,shu} for details),    
\be
\label{pf}
\langle \rho(s,\alpha)\rho(0,0)\rangle_{h}-\langle \rho(s,\alpha)\rangle_{h}
\langle \rho(0,0)\rangle_{h}=R_{2}(s,\alpha)_h=\\ \nonumber
K^2_P(s,0)_h-\left(\frac{d}{ds}K_P(s,\alpha)_h\right)\int_s^{\infty}{K_H}(s',\alpha)_hds'\\ \nonumber
K_P(s,\alpha)_h=\int_{0}^{\infty}dk n_P(k){e^{\alpha^2 u^2}\cos(\pi s k)} \\ \nonumber
K_H(s,\alpha)_h=
\int_{0}^{\infty}dk n_H(k)e^{-\alpha^2k^2}\cos(\pi s k) \nonumber
\ee
where $n_H(k)=1-n_{P}(k)=\frac{1}{1+ze^{k^2/h}}$ is the hole occupation number, 
$z=1/(e^{1/h}-1)$ is the fugacity,
$\alpha$ is the magnetic flux (time), and $h$ is related 
 to $\lambda$ by $ 2\pi h \sim 1/\lambda$ for $h \ll 1$ \cite{prbm}.
 
 The $\alpha \rightarrow 0$ limit in (\ref{pf}) yields  the density-density  
 correlations of the CS model at finite temperature $h \ll 1$ \cite{tsv,ant3} which
 resembles the two point spectral correlations of a disordered
 system at the localization-delocalization transition. The limit 
$h \rightarrow 0$ corresponds with the exact 
 analytical result for the 
 crossover between orthogonal and unitary symmetry
 of a disordered system in the metallic regime \cite{efetov} 
 which is equivalent to the
 exact  dynamical density-density correlations of the CS model \cite{shu}. 
\begin{figure}
\includegraphics[width=0.8\columnwidth,clip]{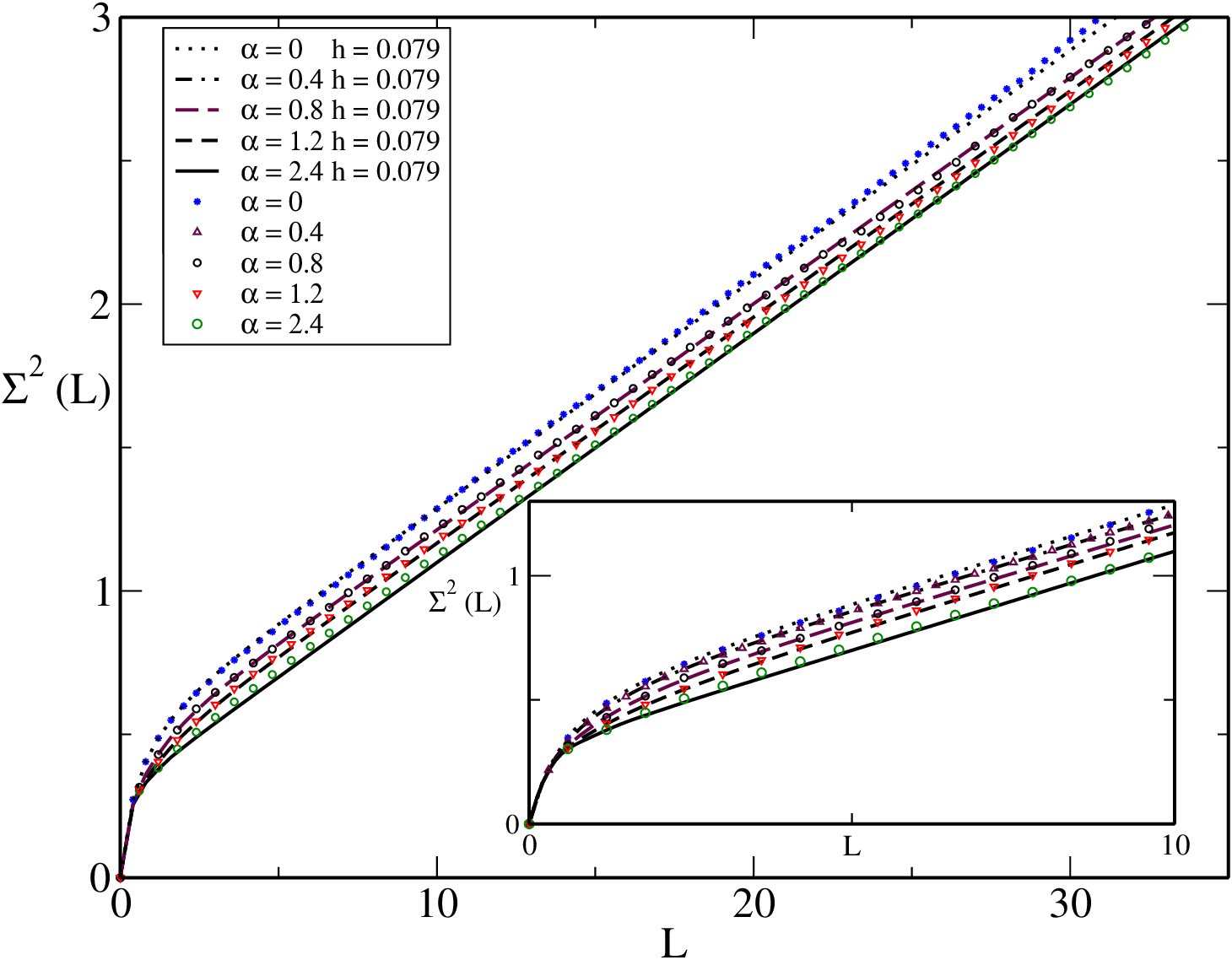}
\caption{Number variance $\Sigma^2(L)$ for weak disorder $\lambda=2$ and different
 flux strengths. Lines are the analytical prediction (\ref{pf}) and symbols
 are the results obtained from the numerical diagonalization of (\ref{mod}) for
  $N=8000$.}    
\label{fig3} 
\end{figure}    
The $s \gg 1, h \ll 1$ limit of (\ref{pf}) also agrees with 
 the density-density 
 correlations obtained in \cite{kawa,ha} by using conformal field techniques.

  Numerical calculations
 also support the validity of our conjecture.
As observed in fig 3, the agreement between the analytical number variance
 based on (\ref{pf}) and the numerical result is  
 excellent for the whole range 
 of fluxes at $\lambda \sim 2$. 
 We remark 
 that no fitting is involved 
 as the value of $h$ utilized corresponds with 
 the analytical prediction $h =\frac{1}{2\pi \lambda} \sim 0.079$ \cite{prbm}.
 In the strong disorder regime (see figure 4) 
 the agreement is also remarkable but in this case
 no relation between $h$ in $\lambda$ is known so   
the parameter $h=0.267$  
 is the best fit to the numerical results. We want to point out that 
 further work is needed to test whether the conjecture (\ref{pf}) is valid beyond
 the $h \ll 1$ limit. 
\begin{figure}
\includegraphics[width=0.8\columnwidth,clip]{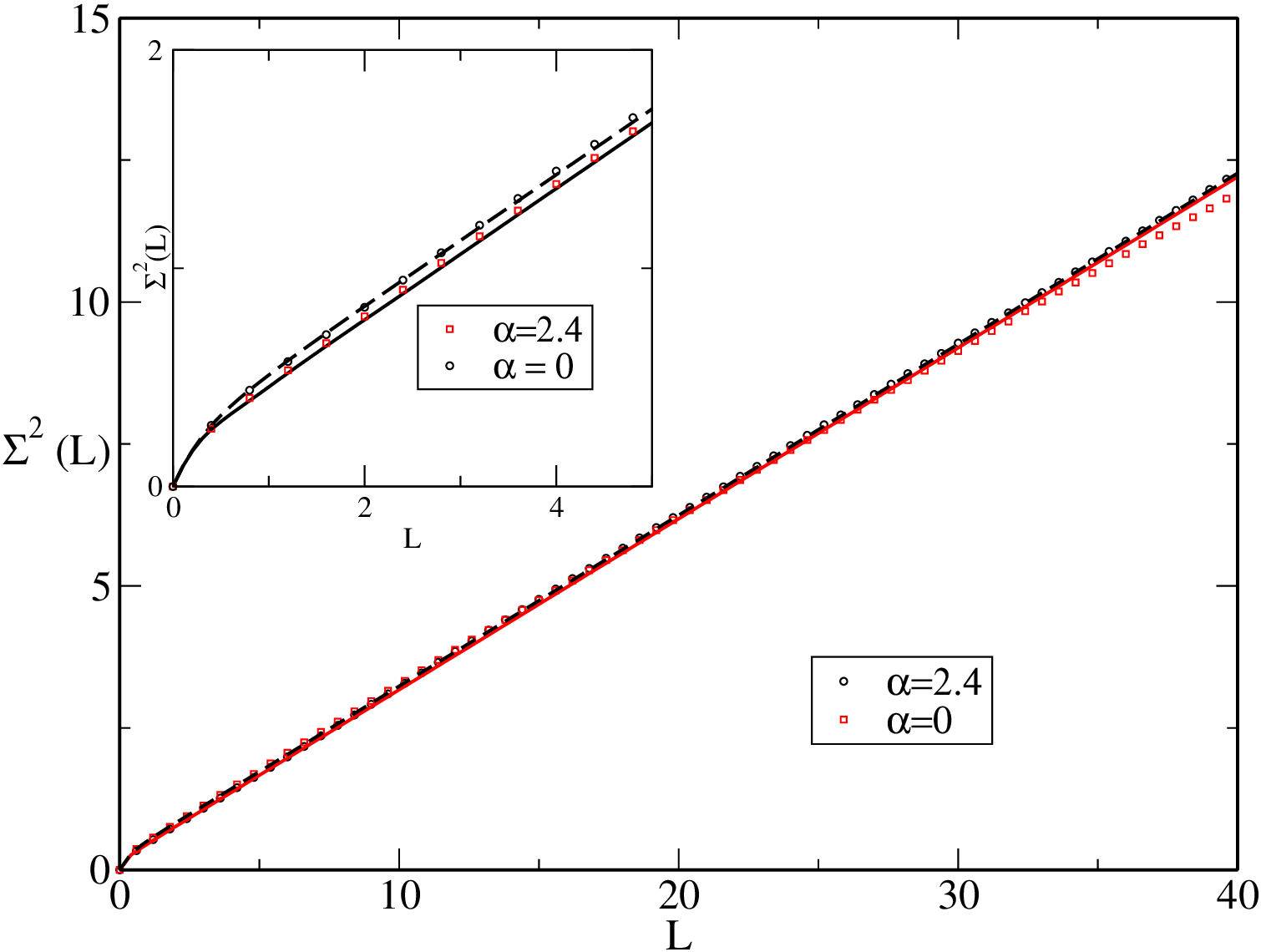}
\caption{Number variance $\Sigma^2(L)$ for a disorder ($\lambda=0.44$) mimicking a true
 3D Anderson transition. The lines represent the analytical prediction (\ref{pf}) ($h=0.267$)
 for no flux (segment) and maximum flux (solid).
  The 
 symbols correspond to the numerical diagonalization of 
 (\ref{mod}) for $N=8000$. The dependence on the flux observed 
 for small $L \sim 1$ (see inset) is also reproduced by (\ref{pf}).}  
\label{fig4} 
\end{figure} 

 Finally we study the relation between 
 classical and quantum properties of our model.  For $\alpha = 0, h \ll 1$,  
 the two level spectral 
 correlation function ($s \gg 1$)  
 can be expressed through the spectral determinant of a
 diffusion operator \cite{andre} describing the classical dynamics 
of (\ref{mod}), 
\vspace{-1mm}
\be
\label{kt}
 R_2(s)&=&-\frac{1}{2\pi^2}\frac{\partial^2
G(s)}{\partial
s^2}+2
\cos(2\pi s)\,e^{2G(s)}
\ee
where,
$e^{G(s)}=\frac{D(s)}{2\pi^2\, s^2}$,$\small{D^{-1}(s)=\prod_{m\neq 0}
\frac{ {\epsilon_m}^{2}+s^2}{{\epsilon_{m}}^2}}\nonumber$
is the spectral determinant 
and $\epsilon_{m}=m/h$ 
are the eigenvalues of the
  (anomalous) diffusion operator.  As expected, the 
($s \gg 1,\alpha = 0$), limit of (\ref{pf}) coincides with (\ref{kt}). 
Unlike short range models, the type of  
 classical diffusion 
 associated with (\ref{mod}) is anomalous \cite{rev}. Such diffusion is  
 described by a fractional Fokker-Planck equation \cite{zas}. 
For band decaying as $1/r^{\gamma}$, $1/2< \gamma < 3/2$, the  resulting 
classical dynamics is superdiffusive with $|r|\sim t^{1/(2\gamma-1)}$\cite{prbm}. In the our case ($\gamma =1$) 
  the associated  fractional Fokker Planck equation is 
 first order in space \cite{fog} and the   
  eigenvalues of the diffusion operator (\ref{kt}) are thus  
 linear in $m$ instead of quadratic as for normal diffusion.
 This link between anomalous classical 
 motion and quantum correlations at the MIT may be utilized to find out 
 conditions for the Anderson transition in quantum chaotic systems. 

In conclusion, we have performed a numerical and analytical investigation
 in a 1D wire with long range disorder and a flux attached to it. 
In the weak 
 disorder regime we have observed that despite the critical character of the model 
 the effect of the flux is still important. 
 By contrast, in the strong disorder limit,   
the spectral correlations are almost independent 
 of the flux except for small eigenvalue separations. We have also conjectured, by exploiting an 
 analogy with the CS model, an exact analytical relation for the
 spectral correlations at the MIT valid for arbitrary flux and disorder $h \ll 1$. We have  suggested that, at least in 1D, the MIT, a quantum mechanic phenomenon,
 may be related with certain features of the classical dynamics (anomalous diffusion). We expect 
 this relation to be relevant in quantum chaos problems.        

Discussions with Gilles Montambaux and Patricio Leboeuf are 
gratefully acknowledged. This work was supported by the European Union
  network ``Mathematical aspects of quantum chaos''.

\end{multicols}

\vspace{-5mm}
\begin{references}
\vspace{-15mm}
\bibitem{wegner80} F.~Wegner, Z. Phys. B {\bf 36},  209 (1980).
\bibitem{janssen} M.~Janssen, Phys. Rep. {\bf 295}, 1 (1998). 
\bibitem{castpel} C.~Castellani and L.~Peliti, J. Phys. A:
Math. Gen. {\bf 19},  L429 (1986);H. Grussbach and M. Schreiber, Phys. Rev.
B {\bf 45} (1993) 6650. 
\bibitem{sko}B.I.~Shklovskii, B.~Shapiro, B.R.~Sears,
P.~Lambrianides and H.B.~Shore, Phys. Rev. B {\bf 47}, (1993) 11487.\
\bibitem{chi} B.L.~Altshuler, I.K.~Zharekeshev, S.A.~Kotochigova and 
B.I.~Shklovskii, JETP 67 (1988) 62.
\bibitem{kravtsov97} V.E.~Kravtsov and K.A.~Muttalib, Phys. Rev. Lett.
{\bf 79}, 1913 (1997).
\bibitem{chalker96} J.T.~Chalker, I.V.~Lerner, and R.A.~Smith,
Phys. Rev. Lett. {\bf 77}, 554 (1996); J.T.~Chalker, V.E.~Kravtsov, and I.V.~Lerner, Pis'ma
Zh. Eksp. Teor. Fiz. {\bf 64}, 355 (1996)
\bibitem{Moshe}M.~Moshe, H.~Neuberger and B.~Shapiro, Phys. Rev. Lett.
{\bf 73}, (1994) 1497; K.A. Muttalib, Y. Chen, M.E.H. Ismail and V.N. Nicopoulos,
Phys. Rev. Lett. {\bf 71}, (1993) 471. 
\bibitem{ant3} A.M.~Garcia-Garcia and  J.J.M.~Verbaarshot, Phys. Rev. E {\bf 67}, 046104 (2003).
\bibitem{tsv}V.E.~Kravtsov and A.M.~Tsvelik,
 Phys. Rev. B {\bf 68}, (2000) 
9888.  
\bibitem{calo}B.~Sutherland, J. Math. Phys. {\bf 12}, 246 (1971);F. Calogero, 
J. Math. Phys. {\bf 10}, 2191 (1969).
\bibitem{anderson}P.W. Anderson, Phys. Rev. {\bf 109}, 1492 (1958).
\bibitem{ono}G.~Yeung and Y.~Oono, Europhys. Lett. {\bf 4}, 1061 (1987). 
\bibitem{yu} C.C.~Yu, Phys.Rev.Lett. {\bf 63}, (1989) 1160.
\bibitem{levitov90} L.S.~Levitov, Phys. Rev. Lett. {\bf 64}, 547 (1990).
\bibitem{prbm} A.D.~Mirlin, Y.V.~Fyodorov, F.-M.~Dittes, J.~Quezada,
and T.H.~Seligman, Phys. Rev. E {\bf 54}, 3221 (1996). 
\bibitem{ever}F.~Evers and A.D.~Mirlin, Phys.Rev.Lett. {\bf 84}, (2000) 3690;E.~Cuevas, M.~Ortuno 
 et.al. Phys. Rev. Lett. {\bf 88}, (2002) 016401;I.~Varga and D.~Braum, Phys. Rev. B {\bf 61}, (2000) R11859.
\bibitem{hof}E. Hofstetter and M. Schreiber, Phys. Rev. Lett. {\bf 73}, (1994) 3137.
\bibitem{bat}M. Batsch, M. Schweitzer et.al  Phys. Rev. Lett. {\bf 77}, (1996)
 1552.
\bibitem{efetov}A.~Altland, S.~Iida, K.B. Efetov, J. Phys. A {\bf 26}, (1993) 3545;
A.~Pandey and M.L.~Mehta, Commun.Math.Phys. {\bf 87}, 449 (1983).
\bibitem{shu}A.~Pandey and P.~Shukla, J.Phys. A, {\bf 24}, (1991) 3907.
\bibitem{ha} Z.N.C. Ha, Phys. Rev. Lett. {\bf 73}, 1574 (1994).  
\bibitem{rev}R.~Metzler and J.~Klafter, Phys. Rep. {\bf 339}, (2000) 1.
\bibitem{kawa}N.~Kawakami and S.~Yang, Phys. Rev. Lett. 67, (1991) 2493.
\bibitem{levitov97} B.L.~Altshuler and L.S.~Levitov, Phys. Rep. {\bf
288}, 487 (1997).
\bibitem{nishi}S.~Nishigaki, Phys. Rev. E {\bf 59}, (1999) 2853
\bibitem{falko}V. I.~Falko and K.B.~Efetov, Phys. Rev. B {\bf 50}, (1994) 11267.
\bibitem{andre} A. V. Andreev and B. L. Altshuler, Phys. Rev. Lett. {\bf 75},
(1995) 902.
\bibitem{zas}G.M.~Zaslavsky, Phys.Rep. {\bf 371}, (2002) 461.
\bibitem{fog}H.C.~Fogedby, Phys. Rev. B {\bf 50}, 1657 (1994).
\end{references}
\end{document}